\documentclass[apj]{emulateapj}
\usepackage{apjfonts}

\def\gax{\mathrel{\raise.3ex\hbox{$>$}\mkern-14mu\lower0.6ex\hbox{$\sim$}}}
\def\lax{\mathrel{\raise.3ex\hbox{$<$}\mkern-14mu\lower0.6ex\hbox{$\sim$}}}
\def\gtorder{\mathrel{\raise.3ex\hbox{$>$}\mkern-14mu
             \lower0.6ex\hbox{$\sim$}}}
\def\ltorder{\mathrel{\raise.3ex\hbox{$<$}\mkern-14mu
             \lower0.6ex\hbox{$\sim$}}}

\begin{document}

\title{The SN~2008S Progenitor Star: Gone or Again Self-Obscured?}

\author{J.~L. Prieto$^{1,7}$, D.~M. Szczygie{\l}$^{2,4}$,
  C.~S. Kochanek$^{2,4}$, K.~Z. Stanek$^{2,4}$,
  T.~A. Thompson$^{2,4}$, J.~F. Beacom$^{2,3,4}$, \newline P.~M. Garnavich$^{5}$, C.~E. Woodward$^{6}$}
 
\altaffiltext{1}{Carnegie Observatories, 813 Santa Barbara St., Pasadena, CA, 91101}
\altaffiltext{2}{Department of Astronomy, The Ohio State University, 140 W. 18th Ave., Columbus OH 43210}
\altaffiltext{3}{Department of Physics, The Ohio State University, 191 W. Woodruff Ave., Columbus OH 43210}
\altaffiltext{4}{Center for Cosmology and AstroParticle Physics, The Ohio State University, 191 W. Woodruff Ave., Columbus OH 43210}
\altaffiltext{5}{University of Notre Dame, 225 Nieuwland Science Hall, Notre Dame, IN 46556}
\altaffiltext{6}{Department of Astronomy, School of Physics and Astronomy, 116 Church Street, S. E., University of Minnesota, Minneapolis, MN 55455}
\altaffiltext{7}{Hubble and Carnegie-Princeton Fellow}

\begin{abstract}

\noindent We obtained late-time optical and near-IR imaging of
SN~2008S with the Large Binocular Telescope (LBT). We find that (1) it
is again invisible at optical ($UBVR$) wavelengths to magnitude limits
of approximately $25$~mag, and (2) while detected in the near-IR
($HK$) at approximately $20$~mag, it is fading rapidly.  The near-IR
detections in March and May 2010 are consistent with dust emission at
a blackbody temperature of $T \simeq 900$~K and a total luminosity of
$L \simeq 40000$~$L_\odot$, comparable to the luminosity of the
obscured progenitor star. If it is a supernova, the near-IR emission
is likely due to shock heated dust since the elapsed time from peak is
too long to support a near-IR dust echo and the decline in luminosity
is shallower than the $^{56}$Co slope. If it is reprocessed emission
from a surviving progenitor, a dust photosphere must have
reestablished itself closer to the star than before the transient
($\sim 40$~AU rather than $150$~AU), unless there is a second, cooler
dust component that dominates at mid-IR wavelengths. The continued
rapid fading at roughly constant temperature favors transient
emission, but the SED peaks in the mid-IR and future Spitzer
observations will be needed to close the case.

\end{abstract}


\section{Introduction}
\label{sec:introduction}

SN~2008S is one of the most mysterious optical transients created by a
massive star in the last decade. It was discovered in February 2008 by
\cite{Arbour2008} in the prolific supernova factory NGC~6946. It was
initially classified as a likely ``supernova impostor" due to its
faint absolute peak magnitude ($M_V \sim -13$~mag) and optical spectra
dominated by narrow Balmer and [Ca~II] lines in emission
(\citealt{Stanishev2008}; \citealt{Steele2008}). NGC~6946 had been
observed by the Large Binocular Telescope (LBT) the previous year, and
the key piece of evidence from these observations was that there was
no optical progenitor \citep{Prieto2008}, which was surprising since
the ``supernova impostors'' are believed to be eruptions from very
massive ($>20$-$30\,M_\odot$), evolved stars (e.g.,
\citealt{Smith2010} and references therein) that should have been
easily visible in the LBT observations.

\begin{table*}
\begin{center}
\caption{LBT magnitudes of SN~2008S \label{tab:magnitudes}}
\begin{tabular}{cccccccc}
\tableline
Date & MJD & $Us$ & $B$ & $V$ & $R$ & $H$ & $K$ \\
(UT) &  & (mag) & (mag) & (mag) & (mag) & (mag) & (mag) \\
\tableline
2008-05-03 & 54589.4  &$21.49\pm0.07$ &$20.86\pm0.03$  &$19.46\pm0.04$ &$18.47\pm0.03$ &$\cdots$       &$\cdots$ \\
2008-05-04 & 54590.4  &$21.52\pm0.08$ &$20.91\pm0.03$  &$\cdots$       &$18.48\pm0.03$ &$\cdots$       &$\cdots$ \\
2008-07-05 & 54652.4  &$22.72\pm0.07$ &$22.27\pm0.03$  &$21.16\pm0.04$ &$20.03\pm0.04$ &$\cdots$       &$\cdots$ \\
2008-11-22 & 54792.1  &$\cdots$       &$23.59\pm0.05$  &$22.50\pm0.05$ &$\cdots$       &$\cdots$       &$\cdots$ \\
2008-11-23 & 54793.1  &$\cdots$       &$23.58\pm0.06$  &$22.56\pm0.05$ &$\cdots$       &$\cdots$       &$\cdots$ \\
2008-11-24 & 54794.1  &$\cdots$       &$23.45\pm0.05$  &$22.45\pm0.05$ &$\cdots$       &$\cdots$       &$\cdots$ \\
2008-11-25 & 54795.1  &$\cdots$       &$23.54\pm0.06$  &$22.60\pm0.05$ &$\cdots$       &$\cdots$       &$\cdots$ \\
2009-03-25 & 54915.5  &$< 24.1$       &$<25.6$         &$<24.8$        &$23.10\pm0.07$ &$\cdots$       &$\cdots$ \\
2009-10-20 & 55124.1  &$<25.2$        &$<25.9$          &$<25.7$       &$< 25.1$       &$\cdots$       &$\cdots$ \\
2009-10-22 & 55126.1  &$<24.9$        &$<25.9$          &$<25.6$       &$< 25.1$       &$\cdots$       &$\cdots$ \\
2009-12-17 & 55182.0  &$\cdots$       &$\cdots$        &$\cdots$       &$\cdots$       &$20.31\pm0.14$ &$\cdots$ \\
2010-03-17 & 55272.5  &$\cdots$       &$\cdots$        &$\cdots$       &$\cdots$       &$<21.4$ &$19.23\pm0.09$  \\
2010-03-18 & 55237.5  &$< 24.6$       &$<25.3$         &$<25.4$        &$<24.9 $       &$\cdots$       &$\cdots$ \\
2010-05-17 & 55333.4  &$\cdots$       &$\cdots$        &$\cdots$       &$\cdots$       &$\cdots$       &$20.27\pm0.15$ \\
\tableline
\end{tabular}
\tablenotetext{}{All the magnitude upper limits are $3\sigma$. The
  estimated start date of the transient is MJD $54485.5 \pm 4$
  \citep{Botticella2009}. $B$, $V$ and $R$ are Bessel filters, $Us$ is
  a high throughput $U$-band interference filter.}
\end{center}
\end{table*}

\begin{figure*}[t]
\centerline{\includegraphics[width=14cm]{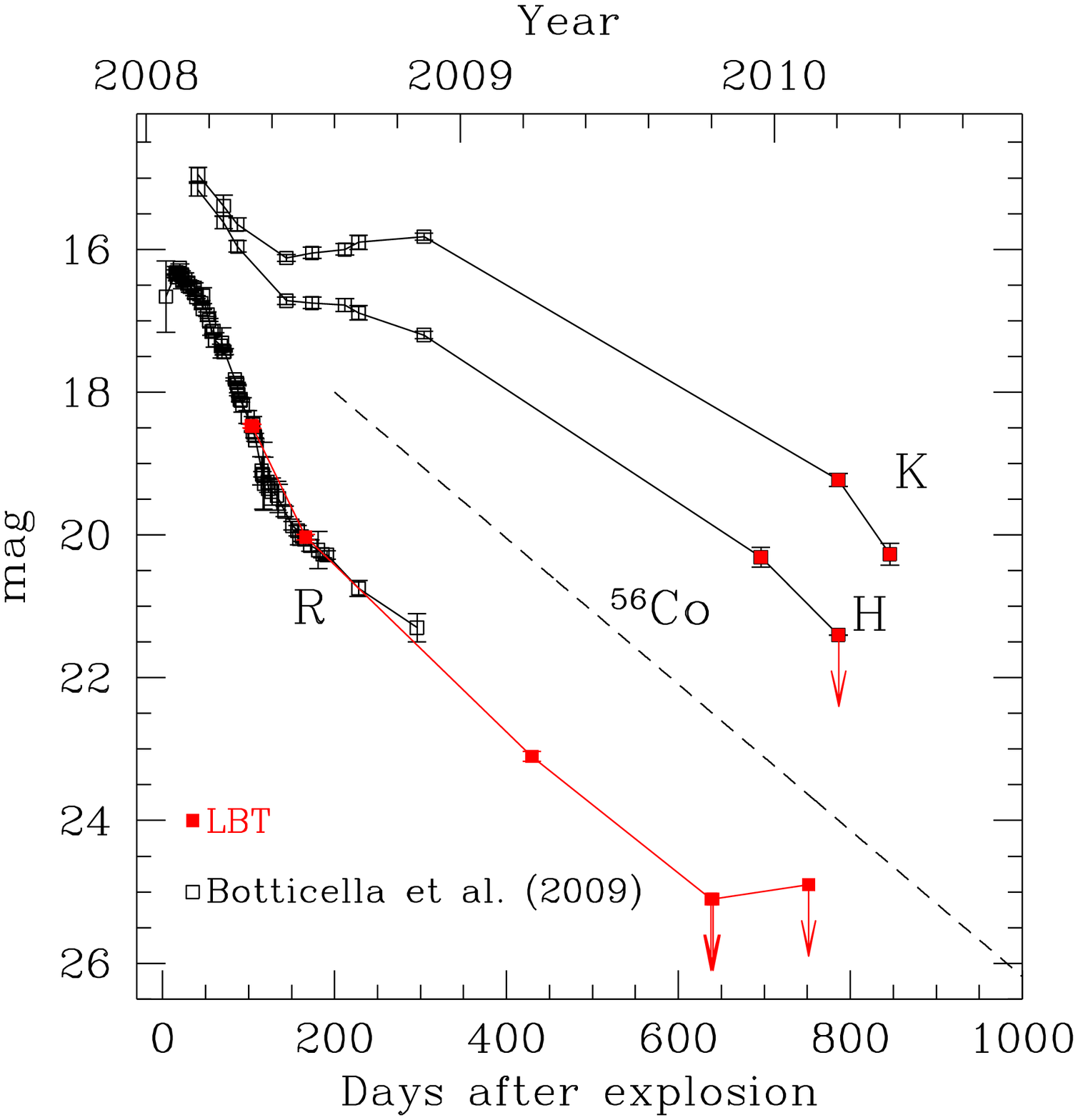}}
\caption{The $R$, $H$ and $K$-band light curves of SN~2008S from
  Botticella et al. (2009, open black points) and the Large Binocular
  Telescope (filled red points). The last $R$ and $H$-band points are
  upper limits.  The dashed line shows the $^{56}$Co decay slope.
  This should properly be compared with the bolometric light curve,
  but this will require Spitzer observations. Botticella et al. (2009)
  found that the bolometric light curve observed after day 120 was
  slightly shallower than the $^{56}$Co decay slope.}
\label{fig:lc}
\end{figure*}

The only means of having an optical eruption from a massive star and
an invisible progenitor is for the star to be self-obscured by dust
that is largely destroyed by the transient.  This possibility was
confirmed when \cite{Prieto2008} found the progenitor star as a $\log
L/L_\odot \simeq 4.5$, $T\simeq 440$~K blackbody in archival Spitzer
data. This luminosity is comparable to that of an evolved $\sim 10\,
M_\odot$ star, and is well below that corresponding to the more
massive stars thought to be required for non-supernova eruptions. 
Subsequent analyses of the progenitor by \cite{Botticella2009} and
\cite{Wesson2010} were consistent with those by \cite{Prieto2008}.

More remarkably, an almost identical event then occurred in NGC~300
(\citealt{Monard2008}).  The progenitor was invisible in the optical
to even tighter limits (\citealt{Berger2008}; \citealt{Bond2009};
\citealt{Berger2009}), but we again found the progenitor as a
self-obscured star of similar luminosity and (dust photosphere)
temperature in Spitzer mid-IR data (\citealt{Prieto2008a};
\citealt{Thompson2009}).  A subsequent analysis of the progenitor by
\cite{Berger2009} agreed with our estimates, and an investigation of
the progenitor based on its neighboring stars by \cite{Gogarten2009}
was consistent with the progenitor being a massive star of order
$10$-$20\,M_\odot$, where the analysis favored the upper portions of
this range but, strictly speaking, the method only provides an upper
mass bound.

In \cite{Thompson2009} we surveyed the galaxy M33 for mid-IR sources
with similar properties to these progenitors and found that they were
astonishingly rare, with only a few such sources in the entire galaxy.
In the mid-IR, these sources have the properties of super-AGB stars,
with properties distinct from other classes of massive stars such as
LBVs and red supergiants. The rarity of these sources compared to all
massive stars, confirmed in our survey of additional galaxies
\citep{Khan2010}, means that the progenitors of the transients are a
very short lived ($\sim 10^4$~years) phase in the evolution of these
massive stars and that there is a causal connection between
obscuration and explosion.

\cite{Thompson2009} concluded that there are a number of possible
mechanisms to explain the nature of these transients and their
progenitors: (1) massive white-dwarf birth; (2) electron-capture
supernova; (3) intrinsically low-luminosity iron core-collapse
supernova; and (4) massive star outbursts. Debates about these
possible origins have been raging ever since then, based both on
theoretical and observational arguments.  They are basically divided
into the (some kind of) supernova camp (\citealt{Prieto2008};
\citealt{Botticella2009}; \citealt{Pumo2009}) and the (some kind of)
massive star outburst camp (\citealt{Berger2009}; \citealt{Smith2009};
\citealt{Bond2009}; \citealt{Kashi2010}).  The outburst camp generally
argues that the progenitor was not a $\sim 10\,M_\odot$ super-AGB star
but a more massive $15-20\,M_\odot$ star (supported by
\citealt{Gogarten2009}), despite their position at the red, high
luminosity end of the AGB sequence in mid-IR color-magnitude diagrams
(\citealt{Thompson2009}; \citealt{Khan2010}) and the low mass compared
to typical stars with LBV outbursts (see \citealt{Smith2010}). The
massive-star outburst interpretation is seriously called into question
by our Spitzer IRS spectrum of the NGC~300 event
(\citealt{Prieto2009}). The mid-IR spectrum resembles that of
carbon-rich proto-planetary nebulae and lacks the silicate-dominated
dust features typical of massive star outbursts (e.g.,
\citealt{Humphreys2006}).  \cite{Wesson2010}, analyzing post-event
Spitzer observations of SN~2008S, also found that the silicate dust
characteristics of high mass stars were inconsistent with the
observations.  \cite{Prieto2009} also note that proto-planetary
nebulae (initial masses $ \ltorder 8\,M_\odot$) have most of the
optical spectral features that led \cite{Smith2009}, \cite{Bond2009}
and \cite{Berger2009} to argue for an outburst from a more massive
($\sim 20\,M_\odot$) star.  Since ``Type IIn'' optical spectroscopic
properties are seen in some proto-planetary nebulae, massive
supergiants, supernova impostors, and the genuine, but very diverse,
Type IIn supernovae, they appear only to be a diagnostic for the
presence of strong interactions between ejecta and a dense
circumstellar medium rather than a diagnostic for the source of the
ejecta.
 
In the end, however, the question is easy to answer -- either the
stars survived, or they did not.  We have been following the SN~2008S
event with the LBT in both the optical and near-IR, and here we report
that the source is again too faint to detect in the optical, and while
detected in the near-IR, it presently is only as luminous as the
progenitor and fading rapidly.  We describe our observations and
results in \S2 and discuss their implications in \S3.

\begin{figure*}
\centerline{\includegraphics[width=18cm]{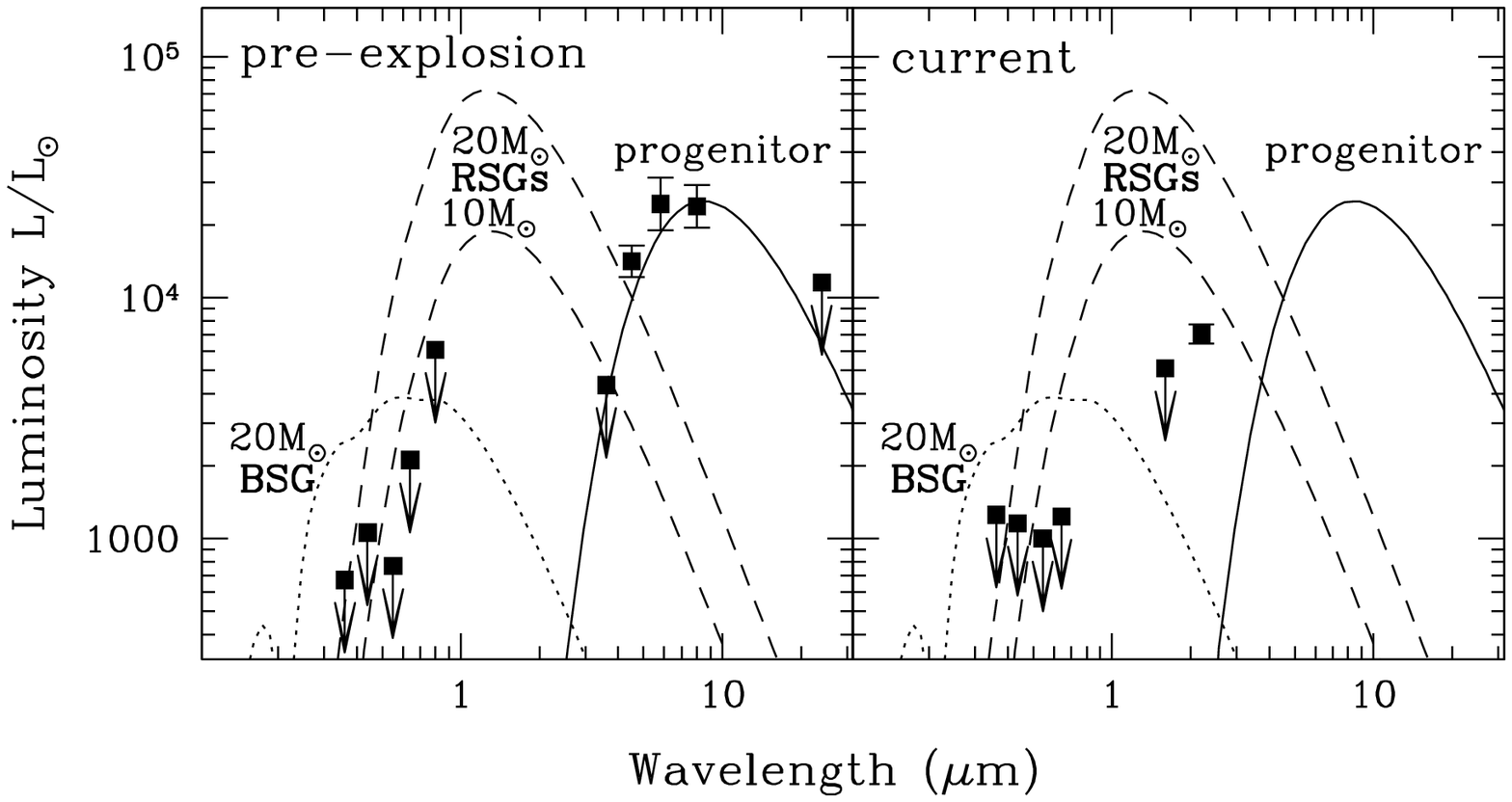}}
\caption{The pre-explosion, progenitor SED (left) and the current SED
  (right) of SN~2008S. The measured magnitudes are converted to
  fluxes, and these are converted to a luminosity as $L=4\pi D^2 \nu
  F_\nu$ where $D=5.6$~Mpc. The SED models are just blackbodies plus
  $A_V=2.13$~mag of total extinction. The $10\,M_\odot$ and
  $20\,M_\odot$ red supergiant models (RSG, dashed curves) are from
  Marigo et al. (2008) and have $T \simeq 3600$ and $3900$~K with
  $\log L/L_\odot = 4.68$ and $5.29$, respectively.  The blue
  supergiant model (dotted curve) is based on SN1987A and has $T
  \simeq 16000$~K and $\log L/L_\odot = 5.0$.  The best fit blackbody
  model (solid curve) for the progenitor has $T=440$~K and $\log
  L/L_\odot=4.54$ \citep{Prieto2008}.}
\label{fig:sed}
\end{figure*}

\section{Observations and Results}

The optical observations were done with the Large Binocular Cameras
(LBC, \citealt{Giallongo2008}), using the LBC/Blue camera for $U$, $B$
and $V$ and the LBC/Red camera for $R$.  The pixel scale of the LBC
cameras is $0\farcs22$.  Since these observations are part of a
program whose overall goal is to use difference imaging to
characterize variable sources, the sub-images obtained for each epoch
were not dithered and SN~2008S was always located at approximately the
same point on Chip~2 of the cameras. Image exposure times were
$300$~sec, generally with two exposures for $U$, $B$ and $V$ and 6
exposures for $R$.  The near-IR observations were made with LUCIFER
(\citealt{Seifert2003}; \citealt{Mandel2008}; \citealt{Ageorges2010})
in the $H$ and $K$ bands using the F3.75 camera with a pixel scale of
$0\farcs12$.  At each dither position we obtained 3 exposures of 33
(10) sec for $H$ ($K$) band.  We obtained 10 on-source and 6
off-source dither positions in a 2-5-2-5-2 off-on-off-on-off pattern,
where the off-source position was shifted 8~arcmin away from the
galaxy.

The optical and near-IR data were reduced using standard methods in
IRAF. The photometry was obtained using DAOPHOT and ALLSTAR
(\citealt{Stetson1987}; \citealt{Stetson1992}).  The optical data was
calibrated using $4-24$ local standards from \cite{Welch2007} for the
$V$ and $R$ bands and from \cite{Botticella2009} for the $U$ and $B$
bands. The near-IR data were calibrated using $3-6$ 2MASS stars in the
field. In both cases we only applied a zero-point offset to convert
the instrumental magnitudes into the standard system. The results are
presented in Table~\ref{tab:magnitudes}, where the magnitude errors
include the uncertainties both in the measurements and in the zero
points. In the cases where we do not detect SN~2008S, we place a
$3\sigma$ upper limit on the magnitude using the standard deviation of
the sky in a region around the source.

Figure~\ref{fig:lc} shows the $H$, $K$ and $R$-band light curves from
Botticella et al. (2009) and our LBT observations, and
Figure~\ref{fig:sed} shows the current SED. The left panel of
Fig.~\ref{fig:sed} shows the constraints on the progenitor's spectral
energy distribution (SED) as compared to typical massive stars. To
make the comparison we used a Galactic plus intrinsic extinction of
$A_V=2.13$~mag \citep{Botticella2009} and the distance of $D=5.6$~Mpc
adopted by \cite{Prieto2008}.  The data points are converted to a
luminosity as $L=4\pi D^2 \nu F_\nu$.  For comparison we show the
extincted SEDs of $10\,M_\odot$ and $20\,M_\odot$ red supergiants
(RSG) using luminosities and effective temperatures from
\cite{Marigo2008}, a $20\,M_\odot$ blue supergiant (BSG) modeled on
SN1987A, and the blackbody that best fit the SN~2008S progenitor data.

In the optical ($UBVR$), the source is again too faint to correspond
to a massive ($>10 \,M_\odot$) evolved star, with limits on its
brightness similar to those for the progenitor (see right panel in
Fig.~\ref{fig:sed}). The extinction would have to be increased from
the $A_V\simeq 2.1$~mag estimated to be present post-explosion
(\citealt{Botticella2009}) to $A_V\sim 3.6-5.8$~mag in order to
obscure the models shown in Fig.~\ref{fig:sed}. The transient is still
detectable in the near-IR, but it is fading rapidly with a slope of
approximately $6 \pm 1$~mag/year at $K$ band that is significantly
steeper than the mean slope of $2.3 \pm 0.1$~mag/year between the late
phases of the \cite{Botticella2009} light curve and our first LBT
observation.  The SED is rising to the red with $H-K > 2.2$~mag.  If
we extrapolate the H-band flux from December 2009 to March 2010 using
the slope of the $K$-band light curve, we estimate $H \simeq 21.9$~mag
and thus $H-K \simeq 2.7$~mag, which is significantly redder than the
$H-K \simeq 1.4$~mag color in the late phases of
\cite{Botticella2009}.

We can roughly estimate a temperature and luminosity for the March
2010 epoch.  Fitting a blackbody to the measured $K$-band flux and
either the $H$-band magnitude limit ($20.4$~mag) or the extrapolated
estimate ($20.9$~mag), we get temperatures of $T \simeq 900$~K and
750~K and luminosities of $L\simeq 68000\,L_\odot$ and
$130000\,L_\odot$, respectively.  With a $\lambda^{-1}$ emissivity
law, the estimated temperatures and luminosities are lower, with $T
\simeq 800$~K and $700$~K and $L \simeq 50000\,L_\odot$ and $95000\,
L_\odot$.  With the further fading between March and May 2010, the
source luminosity is now comparable to the estimated luminosity $L
\simeq 40000\,L_\odot$ of the progenitor star (\citealt{Prieto2008};
\citealt{Botticella2009}; \citealt{Wesson2010}).

\section{Discussion}

\cite{Thompson2009} proposed that SN~2008S and the NGC~300 transient
were the archetypes of a new class of transients potentially including
the M85 OT-1 transient (\citealt{Kulkarni2007};
\citealt{Pastorello2007}), SN~1999bw (\citealt{Li2002} and references
therein), and now PTF10fqs (\citealt{Kasliwal2010}). The initial
defining characteristics were (1) a dust-enshrouded progenitor without
optical counterpart and mid-IR magnitudes that places them at the tip
of the AGB sequence in a mid-IR CMD, and (2) a low-luminosity
transient ($-13 \gtrsim M_{V} \gtrsim -15$) with narrow lines in
emission in the spectra ($v \lesssim 3000$~km/s), and signs of a
circumstellar dust excess at near-IR and mid-IR
wavelengths. Examinations of the dust properties
(\citealt{Prieto2009}; \citealt{Wesson2010}) suggest (3) that the dust
is carbonaceous rather than the silicate dust seen in massive stars.

Here we add (4) that the progenitor either does not survive or must
return to its dust enshrouded state. As the right panel of
Fig.~\ref{fig:sed} shows, the LBT data already rule out the presence
of a massive, evolved star unless it has reconstituted an optically
thick, dusty envelope. The present optical limits are somewhat
stronger than those for the progenitor, and the near-IR detections
already rule out RSGs more massive than $10\,M_\odot$. The total
luminosity is now comparable to that of the progenitor and emerges
mainly in the mid-IR, but it is also continuing to rapidly fade.
Recently, \cite{Ohsawa2010} presented late-time AKARI mid-IR
observations of the NGC~300 transient that show the transient is again
self-enshrouded with an SED that peaks at $\sim 3-4$~$\mu$m ($T \sim
600$~K) and total bolometric luminosity $\sim 5$ times the luminosity
of the mid-IR progenitor.

Let us first consider the possibility that the emission is again due
to the progenitor. With roughly the same luminosity but double the
temperature, the dust photosphere must be four times closer to the
star, at $R_{BB} \simeq 40$~AU~\footnote{There are differences in the
  sizes quoted for the dust around SN~2008S. \cite{Prieto2008} assume
  an infinite wind and estimate a photospheric radius of $150$~AU,
  while \cite{Botticella2009} and \cite{Wesson2010} generally discuss
  the geometric boundaries of the dust distribution.}, although we
can't rule out the presence of a cooler dust component that dominates
the bolometric luminosity and peaks in the mid-IR. If the optical
depth is due to a constant velocity wind, this in turn requires
decreasing the mass loss rate $\dot{M}$ (or opacity per unit mass) or
increasing the wind velocity $v_w$ by a factor of 4 relative to the
progenitor.  Producing the near-IR time variability is difficult in
this scenario because the characteristic time scale $R_{BB}/v_w$ is
$\sim 2$~years for $v_w \simeq 100$~km/s while the $K$ band flux
changed by over a factor of 2 in only 60 days. Thus, it seems unlikely
that the system has returned to its pre-transient state.

The rapid fading strongly suggests that the present emission is a
continuation of the transient.  Since we are now 800~days post
explosion, the emission can no longer be explained as an infrared
echo. At this point, echos from the transient peak are produced from a
minimum distance of $ct/2 \simeq 70000$~AU, and this is simply too
distant for dust heated by a transient with a peak luminosity of order
$10^7 \,L_\odot$ \citep{Botticella2009} to produce significant near-IR
emission. While the near-IR emission has roughly decayed at the rate
expected from $^{56}$Co decay (1.023~mag/100 days, see
Fig.~\ref{fig:lc}) the drop in the estimated bolometric luminosity is
significantly slower, and it is unclear how the positron heating could
be efficiently converted to near-IR emission. The last possibility is
shock heating of pre-existing dust. \cite{Botticella2009} and
\cite{Wesson2010} estimate that dust survived outside of
$1000-2000$~AU, which would be reached after 800~days by a shock
moving at $2000-4000$~km/s. Such high velocities were observed in
some early line components (\citealt{Botticella2009}). For the heavy
$\dot{M} \sim 10^{-4}\,M_\odot$/year wind believed to have surrounded
the progenitor, the shock luminosity of $(1/2) \dot{M} v_s^3/ v_w \sim
10^6 \,L_\odot$ for $v_w \sim 50$~km/s and $v_s \sim 3000$~km/s, is on
the order of what is needed to produce the near-IR emission with $\sim
10\%$ efficiency of emission by shocked dust (\citealt{Draine1981}).
However, this seems a stretch given the time and velocity scales, and
it would be simpler to use dust forming in the shocked material as
advocated by \cite{Botticella2009}.

At its present rate of fading in the near-IR, SN~2008S will
effectively be invisible to ground based observatories when it next
rises, and finally closing the case will need a combination of HST
and, more importantly, Spitzer observations that will be obtained over
the next year.  The HST observations can detect or rule out the
presence of a star in the near-IR to significantly deeper limits than
possible from the ground due to both its sensitivity and resolution.
With two epochs of data showing some variability, the source can be
unambiguously identified even if very faint. The Spitzer observations
will accurately determine the temperature and luminosity of the
source. If it continues to decay as rapidly as we observed in the
near-IR, it should be significantly fainter than the progenitor star
in 2011.

These late time observations will be crucial to understanding this new
class of transient sources, particularly since it is also possible for
the survivor to be fainter than the progenitor in several of the
possible scenarios.  It could be subluminous as a result of the
outburst and then will slowly return to thermal equilibrium
(\citealt{Smith2009}). Or, as suggested by \cite{Thompson2009} and
discussed more fully in \cite{Prieto2009}, if SN 2008S was the
explosive birth of a massive white dwarf, we would expect the
bolometric luminosity to approach nearly Eddington for a $\sim
1M_\odot$ object, $\sim 3\times 10^4 L_\odot$.

\acknowledgements 

We thank G. Cresci, J. Hill, R. Humphreys, and A. Quirrenbach for
suggestions and comments. Based in part on observations made with the
Large Binocular Telescope. The LBT is an international collaboration
among institutions in the United States, Italy and Germany. The LBT
Corporation partners are: the University of Arizona on behalf of the
Arizona university system; the Istituto Nazionale di Astrofisica,
Italy; the LBT Beteiligungsgesellschaft, Germany, representing the Max
Planck Society, the Astrophysical Institute Potsdam, and Heidelberg
University; the Ohio State University; and the Research Corporation,
on behalf of the University of Notre Dame, University of Minnesota and
University of Virginia. JLP acknowledges support from NASA through
Hubble Fellowship grant HF-51261.01-A awarded by STScI, which is
operated by AURA, Inc. for NASA, under contract NAS~5-2655. CSK, JLP,
KZS, DS and TAT are supported in part by NSF grant AST-0908816. JFB is
supported by NSF CAREER Grant PHY-0547102.

{\it Facilities:} \facility{LBT}

\end{document}